\documentclass[aps,prb,twocolumn,10pt]{revtex4-1}
\usepackage[latin9]{inputenc}
\usepackage{amsmath}
\usepackage{amssymb}
\usepackage{graphicx}
\usepackage{esint}
\usepackage{epstopdf}

\usepackage[usenames,dvipsnames]{color} 

\newcommand{\red}[1]{{\color{black} #1}}



\begin{document}

\title{Regularized quasinormal modes for  plasmonic resonators and open cavities}

\author{Mohsen {Kamandar Dezfouli}}
\email{m.kamandar@queensu.ca}
\affiliation{Department of Physics, Engineering Physics and Astronomy, Queen's
University, Kingston, Ontario, Canada K7L 3N6}
\author{Stephen Hughes}
\email{shughes@queensu.ca}
\affiliation{Department of Physics, Engineering Physics and Astronomy, Queen's
University, Kingston, Ontario, Canada K7L 3N6}
\begin{abstract}
Cavity mode theory and analysis of open cavities and plasmonic particles is an essential component of
optical resonator physics, offering considerable insight  and efficiency for connecting to classical and quantum optical properties such as the Purcell effect.
However, obtaining the dissipative modes in normalized form for arbitrarily shaped open cavity systems is notoriously difficult, often  involving complex spatial integrations, even after performing the necessary full space solutions to  Maxwell's equations. The formal solutions are termed   quasinormal modes  which are known to diverge in space, and additional techniques are frequently required to obtain more accurate field representations in the far field. In this work we introduce a new finite-difference time-domain technique that can obtain normalized quasinormal modes using a simple dipole-excitation source and an inverse Green function technique, in real frequency space, without having to perform any spatial integrations. Moreover, we show how these modes are naturally regularized  to ensure the correct field decay behaviour in the far field, and thus can be used at any position within and outside the resonator. We term these modes ``regularized quasinormal modes'' and show the reliability and generality of the theory, by  studying the generalized Purcell factor of quantum dipole emitters near metallic nanoresonators, hybrid devices with metal nanoparticles coupled to dielectric waveguides, as well as  coupled cavity-waveguides in photonic crystals slabs. We also directly compare our results with full-dipole simulations of Maxwell's equations without any approximations and show excellent agreement.

\end{abstract}
\maketitle

\section{Introduction}

Over the past few decades, optical cavity structures have been used for an incredibly wide range of photonic applications \cite{Vahala2003}.
In many of these cavity devices, exploiting a modal description of the optical system
is of great benefit, not only for providing good physical intuition,
but also because of the significant efficiency that it brings to the theoretical
investigations of a typical problem. Moreover, in quantum optics,  a modal description is  a requirement for second quantization. 
However, in contrast to \textit{normal
modes} (with real eigenfrequencies), that are  suited for describing closed optical systems (and
sometimes very low loss systems) where energy losses are not
a major concern, for dissipative optical cavities in general, the open cavity
\textit{quasinormal modes} (QNMs) must be employed. 
This is particularly true for low quality factor ($Q$) resonators, though also for high-$Q$ structures such as photonic crystal cavities
\cite{Kristensen2012}. Low-$Q$ broadband resonators are also now widely exploited
in plasmonic devices, which have been proposed and used
for applications such as sensing 
\cite{Kneipp1996,Nie1997,Zhang2013,Yampolsky2014,Chikkaraddy2016},
hybrid integrated photonics 
\cite{Barth2010,Mukherjee2011,hybrid2012,hybrid2012_2,hybrid2013,Castro-Lopez2015,Espinosa-Soria2016,Chai2016,Doeleman2016,KamandarDezfouli2017,Ortuno2017}
and broadband single photon sources \cite{Chang2006,Esteban2010,Russell2012,Belacel2013,Akselrod2014,Lu2014,Ruesink2015,Baaske2016,Koenderink2017}.
For any material system,  the open-cavity QNMs are associated
with complex eigenfrequencies whose imaginary parts quantify the system
losses, and thus they require a more generalized normalization \cite{Leung1994,Kristensen2012,Sauvan2013,Kristensen2015,Bai2013}
beyond the standard Hermitian theories. In recent years,  cavity QNMs have been successfully
used to calculate various optical quantities of interest, such as
the generalized effective mode volume \cite{Leung1994,Kristensen2012}, enhanced
spontaneous emission factor of dipole emitters, and  the electron
energy-loss spectroscopy (EELS) maps for plasmonic resonators
\cite{Abajo2008,Ge2016}.

Calculation and normalization of QNMs has been 
demonstrated using both time-domain techniques, such as finite-difference time-domain (FDTD), and frequency-domain methods (such as finite-element solvers) \cite{Kristensen2012,Ge2014,Sauvan2013,Bai2013}. However, all of these approaches are somewhat specialized
and all have pros and cons. Also, most of these computational methods require a complex spatial integration of the optical fields, involving volume and
surface integrals \cite{Kristensen2012,Kristensen2015,Leung1996JOSAB,Muljarov2010}, or coordinate transforms that exploit
PMLs (perfectly matched layers) as part of the cavity structure \cite{Sauvan2013}.
Therefore, implementation of such QNM normalization approaches is often quite involved and, for this reason, it is still quite common
 to compute cavity modes in a rather ambiguous way, often treating
 the  mode of interest as a normal mode (in some finite computational domain), or as the solution to the scattering problem (which is obviously not a true mode). 
Moreover, in structures with a background periodic index, such as a cavity coupled to a photonic crystal waveguide, the spatial normalization approaches generally fail, and a more careful regularization is required \cite{KristensenOL2014,Malhotra2016}.
Thus there is  a need to develop simpler computational approaches to obtain the QNMs without the need for such a complicated integration procedure.

Recently, using a frequency domain approach, an intuitive dipole normalization technique was developed by Bai {\it et al.}~\cite{Bai2013}, implemented using COMSOL \cite{comsol}, where the self-consistent
response to a dipole excitation is used to obtain an {\it integration-free}
normalization for the QNM. In this approach, a search in frequency space is first performed to identify the QNM resonant frequency (pole) and then an additional simulation is performed very near the  resonance frequency to capture the dominant cavity mode of interest. This method relies on the ability of the frequency-domain
solvers to locate a frequency pole of the resonance in complex frequency space. Such an approach can be highly accurate and insightful, but requires implementation in a frequency domain solver, which can often require significant amounts of memory for larger device simulations such as coupled resonator-waveguide systems, and even single resonator structures \cite{Akselrod2016}. Additional frequency domain Maxwell's solvers have since been developed for obtaining QNMs of plasmonic devices \cite{MNPBEM} that can also be used to perform QNM calculations \cite{MNPBEM_SciRep}, which is cited to work  best for nanoparticles of a certain size \cite{MNPBEM}. \red{Very recently, a generalization approach to normal mode expansion of the Green fucntion for lossy resonators has been introduced \cite{Bergman}, where modes are defined with permittivity rather than frequency. While the technique is quite general and insightful, it  provides single-frequency information and can also suffer from already discussed memory requirement issues (the implementation has been done in COMSOL) for arbitrary sized devices.}

\red{In this work, we describe a new integration-free method for obtaining normalized QNMs of arbitrary open cavities, which uses a simple dipole source excitation technique in real frequency space. We also describe and show how this new technique calculates ``regularized'' QNMs \cite{Ge2014njp}, which ensures well-behaved (non-divergent) fields far away from the resonator, without the need for including a sum over many modes or by carrying our formal regularization through a Dyson integration approach---which reconstructs the regularized fields outside the resonator using  the QNM fields inside \cite{Ge2014njp}. Although our approach is quite general, we will implement it using FDTD (from Lumerical Solutions \cite{lumerical}) that is arguably one of the most general and well used computational techniques in nanoplasmonics \cite{Maier2006,Schuller2010,Ameling2013,Eter2014,Akselrod2016}, and
in nanophotonics in general \cite{fdtdref3,fdtdref1,fdtdref2,fdtdref4}. In Section \ref{sec:Theory}, we introduce simple transparent formulas that can be used
by a general user of any FDTD software (or similar time-dependent Maxwell solver) to compute the system QNMs  in an easy-to-use and reliable manner.} In Section
\ref{sec:Results}, as an application of this technique,
confirmed with full vectorial dipole calculations of Maxwell's equations,
we study the spontaneous emission enhancement (generalized Purcell factor) of a  dipole
emitter placed nearby different plasmonic systems including hybrid systems of metals and dielectrics, as well as coupled cavity-waveguide devices in photonic crystal slabs;  these resonator systems are motivated by recent practical interest in design of different devices such as transmission line filters and single photon sources 
\cite{hybrid2012,hybrid2012_2,hybrid2013,Castro-Lopez2015,Espinosa-Soria2016,Axelrod2017,Malhotra2016}. In \ref{subsec:metals}, we  explore the gold cuboid and dimer structures, to confirm the accuracy of the technique over a wide range of spatial positions and frequencies, where we also comment on how efficient our new technique is compared to existing methods. In \ref{subsec:reg}, we then
explore the regularized nature of our newly developed technique in obtaining the far-field optical response that eliminates  concerns regarding the divergent behavior of the QNMs, e.g., for use with studies such as transmission and scattering into the far field. In addition, we also show excellent agreement with a previous FDTD QNM approach that uses a spatial integration approach with filtering \cite{Ge2014}, and point out several clear advantages of the current approach (including a drastic increase in efficiency, and no spatial integration at all for normalization or far field regularization).
In \ref{subsec:hybrids}, we extend the applicability and reliability of the technique to hybrid devices where plasmonic resonators are coupled to periodic dielectric waveguides. Finally, in \ref{subsec:PC}, we also provide some examples of dielectric cavities, including a complicated coupled cavity-waveguide design in a photonic crystal slab. This later is a particularly  hard problem for obtaining the QNMs (of the coupled system), but is shown to be easily computed with the current technique. In section \ref{sec:Conclusions}, we present our conclusions.

\begin{figure}
\begin{center}
\includegraphics[width=.95\columnwidth]{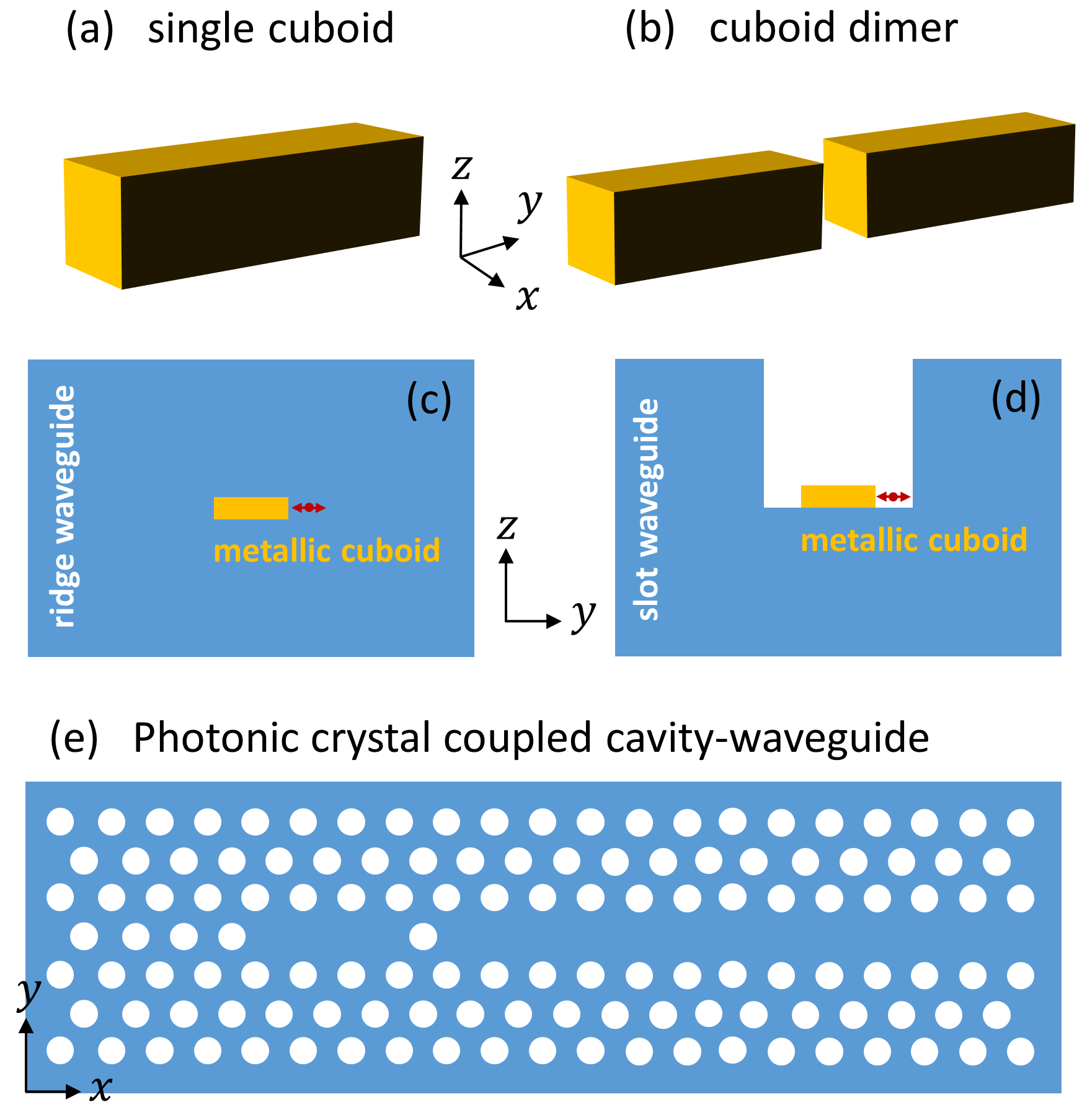}
\caption{Examples of plasmonic, hybrid metal-dielectric and photonic crystal cavity devices for use in various nanophotonics applications, all of which we will study in this paper. (a-b) show plasmonic single and dimer cuboids placed in a homogeneous medium. (c-d) show hybrid devices where the single cuboid is either
embedded inside a ridge waveguide or placed on top of a slot waveguide. The red double arrows show the location and orientation of dipole excitation used in this work where a 2D projection through cuboid center is used. (e) A top down view of a 3D coupled cavity-waveguide device using photonic crystal slabs for single photon source applications. \label{fig:schematic}}
\end{center}
\end{figure}

\section{Theory\label{sec:Theory}}

In this section, we present our  main technique for calculating the regularized QNMs. While our approach can be expanded for computing several QNMs, we focus on a single mode picture, as it is quite often the most desired case, e.g., for  applications with single photon sources and lasing. While all time-domain techniques can become problematic
for computing closely overlapping QNMs (in frequency), there are no additional limitations beyond those that are well known to most time domain Maxwell solvers. 

\subsection{Existing QNM theory\label{sec:ExistingTheory}}
In general, any open cavity system can support several QNMs, $\tilde{\mathbf{f}}_{{\mu}}\left(\mathbf{r}\right)$, over the frequency band of interest. These QNMs can be defined as the  solutions to the Helmholtz equation,
\begin{equation}
\boldsymbol{\nabla}\times\boldsymbol{\nabla}\times\tilde{\mathbf{f}}_{{\mu}}\left(\mathbf{r}\right)-\left(\dfrac{\tilde{\omega}_{{\mu}}}{c}\right)^{2}\varepsilon\left(\mathbf{r},\omega\right)\,\tilde{\mathbf{f}}_{{\mu}}\left(\mathbf{r}\right)=0,
\end{equation}
with open boundary conditions, through the Silver-M\"uller radiation condition \cite{Kristensen2015}. Here, $\varepsilon\left(\mathbf{r},\omega\right)$ is the 
dielectric function (possibly complex) of the system and $\tilde{\omega}_{{\mu}}=\omega_{{\mu}}-i\gamma_{{\mu}}$
is the complex resonance frequency that can also be used to quantify the
system quality factor as $Q_{\mu}=\omega_{\mu}/2\gamma_{\mu}$. These QNMs, once normalized, can be used to construct the transverse Green function through 
\cite{Leung1994,Ge2014njp}
\begin{equation}
\mathbf{G}\left(\mathbf{r},\mathbf{r}_{0};\omega\right)= \sum_{\mu} A_{\mu}\left(\omega\right)\,\tilde{\mathbf{f}}_{\mu}\left(\mathbf{r}\right)\tilde{\mathbf{f}}_{\mu}\left(\mathbf{r}_{0}\right),\label{eq:GFwithSUM}
\end{equation}
 at all frequencies around
the mode and at all locations nearby the scattering geometry, \red{where the QNMs can form a complete basis \cite{Leung1996JOSAB,Leung_PRA_Completeness}. For simplicity, we have defined the spectral function} $A_{\mu}(\omega)=\omega^{2}/2\,\tilde{\omega}_{\mu}\left(\tilde{\omega}_{\mu}-\omega\right)$.
Considering  a single QNM, $\tilde{\mathbf{f}}_{\rm c}(\mathbf{r})$, the single mode Green function can be written as
\begin{equation}
\mathbf{G}_{\rm c}\left(\mathbf{r},\mathbf{r}_{0};\omega\right) \approx A_{\rm c}(\omega)\,\tilde{\mathbf{f}}_{\rm c}\left(\mathbf{r}\right)\tilde{\mathbf{f}}_{\rm c}\left(\mathbf{r}_{0}\right),
\label{eq:GF_QNM}
\end{equation}
where again this strictly holds only nearby the ``cavity region'' (typically at distances where a quantum emitter still feels a Purcell factor enhancement) and diverges at locations far away. \red{The exact position for which the divergent behavior of the QNM appears, depends on the quality factor of the open cavity under investigation. For low quality factors such as those in plasmonics, the divergent behavior can appear at around a few microns away from the resonator (which will be farther away for higher quality factors such as those in photonic crystal cavities).} In any case, the divergent behavior is clearly unphysical for real fields, as we know there should be no enhanced emission in the far field, and the total field must be convergent.  This problem can be partly avoided (or fixed) by employing a Dyson equation formalism to reconstruct the full Green function at locations away (outside) from the cavity region \cite{Ge2014njp}. \red{The idea behind the Dyson equation is to self-consistently obtain the solution to the scattering geometry at all locations using the preexisting knowledge of the background Green function
and field solution inside the scattering geometry \cite{OliverMartinDysonGF}. Using an accurate QNM solution within the cavity region, one can obtain a ``regularized'' mode from,
\begin{align}
\tilde{\mathbf{F}}_{\rm c}(\mathbf{R}) = \int_{\rm cavity} \mathbf{G}_0(\mathbf{R},\mathbf{r^{\prime}};\omega)\,\Delta \varepsilon(\mathbf{r^{\prime}},\omega)\cdot\tilde{\mathbf{f}}_{\rm c}(\mathbf{r^{\prime}})\,{\rm d} \mathbf{r}',\label{eq:bigF}
\end{align}
for any position ${\bf R}$ outside the resonator. Here, $\mathbf{G}_0$ is the Green function for the background medium and $\Delta\varepsilon({\bf r},\omega)=\varepsilon({\bf r},\omega)-\varepsilon_B(\bf r)$ is the total dielectric constant minus the background term $\varepsilon_B(\bf r)$. In the case of isolated resonators, $\mathbf{G}_0$ can be considered as the (analytically known) homogeneous space Green function \cite{Novotny}, and for resonators coupled to waveguides, could be the background waveguide Green function~\cite{Kristensen:17}. Notably, the only assumption used to derive this expression is the validity of the single QNM description within the resonator.} Indeed, one can also use this expression in complex frequency space to self-consistently obtain the divergent QNM outside the resonator:
\begin{align}
\tilde{\mathbf{f}}_{\rm c}(\mathbf{R}) = \int_{\rm cavity} \mathbf{G}_0(\mathbf{R},\mathbf{r^{\prime}};\tilde\omega)\,\Delta \varepsilon(\mathbf{r^{\prime}},\tilde\omega)\cdot\tilde{\mathbf{f}}_{\rm c}(\mathbf{r^{\prime}})\,{\rm d} \mathbf{r}'.
\label{eq:bigFcmplx}
\end{align}
In any case, the regularized QNM, $\tilde{\mathbf{F}}_{\rm c}(\mathbf{r})$ can be used in a similar Green function expansion as in Eq.\,\eqref{eq:GF_QNM} to obtain physically meaningful quantities far outside the resonator, where it is fully expected that a single QNM approach will breakdown.
This ``regularized'' mode can then be used at all positions outside the  resonator, and has previously been shown to be highly accurate when compared to full dipole calculations \cite{Ge2014njp}. Thus, the general goal with practical QNM theory is to obtain $\tilde{\bf f}_{\rm c}$ and then ${\tilde{\bf F}}_{\rm c}$; however, this now requires two complex integrations, one to first obtain the normalized QNM, and  one to obtain the regularized QNM; especially with metallic resonators, this additional integration can  be a complicated process (typically using nm-size grids) and indeed particularly problematic when the background is not so well defined (e.g., in the case of a resonator coupled to an infinite waveguide).

\subsection{Integration-free QNM calculation\label{sec:NewTechnique}}
We now describe our new integration-free dipole technique for accurately obtaining the QNM, and we also show how it naturally provides the regularized QNM in the far field without the need for further treatment. \red{As discussed before, QNMs form a complete basis only nearby the resonator and result in divergent fields in the far-field region. In general, to assure the correct behavior when propagating to the far-field, it is essential to involve all other system modes into our Green function expansion. This may include all QNMs in complex frequency space as well as the homogeneous medium propagating fields in real frequency space. Mathematically this can be written as $\mathbf{G} = \mathbf{G}_{\rm c} + \mathbf{G}_{\rm others}$ where $ \mathbf{G}_{\rm others}$ accounts for all necessary additions to  Eq.\,\eqref{eq:GF_QNM} in the far-field. Indeed, the Dyson regularization technique discussed above self-consistently includes such effects.} In contrast, we will adopt an even simpler approach to regularization; we introduce the following ansatz:
\begin{align}
\mathbf{G}\left(\mathbf{r},\mathbf{r}_{0};\omega\right) \equiv 
A_{\rm c}\left(\omega\right)\,\tilde{\mathbf{f}}^{\rm r}_{\rm c}\left(\mathbf{r}\right)
\tilde{\mathbf{f}}^{\rm r}_{\rm c}\left(\mathbf{r}_{0}\right),\label{eq:GF_rQNM}
\end{align}
where $\tilde{\mathbf{f}}^{\rm r}_{\rm c}(\mathbf{r})$ is the real-frequency  obtained QNM that is now also regularized in the far-field, and thus we call it a renormalized QNM (rQNM) to distinguish it from the usual far field behavior from a (spatially) divergent QNM. This means that, even in the far-field, a single mode expansion (using a rQNM) can be used to obtain physically meaningful quantities. Note that the subscript ``${\rm c}$'' on the Green function is no longer needed as we assume that the Green function of Eq.\,\eqref{eq:GF_rQNM} is an accurate representation of full system Green function at all positions, at frequencies close to the resonance frequency $\omega=\omega_{\rm c}$. 

Consider now a point-source simulation of Maxwell's equation at location
$\mathbf{r}_{0}$, that can be used to obtain the system response
at any location, also returning the dipole self-response term, $\mathbf{G}\left(\mathbf{r}_{0},\mathbf{r}_{0};\omega\right)$.
This is achieved by monitoring all three components of the electric
field at the dipole location to obtain the numerical Green function:
\begin{equation}
G_{ij}\left(\mathbf{r}_{0},\mathbf{r}_{0};\omega\right)=\frac{E_{i}\left(\mathbf{r}_{0},\omega\right)}{P_{j}\left(\omega\right)},\label{eq:GFnumeric}
\end{equation}
where $E_{i}\left(\mathbf{r}_{0},\omega\right)$ is the $i$th component
of the monitored electric field and $P_{j}\left(\omega\right)$ is dipole
polarization introduced along the $j$th direction, with $i$ and
$j$ representing Cartesian coordinates. Assuming that a relatively accurate estimation
of the complex frequency for the localized resonance is available, the ansatz of Eq.~\eqref{eq:GF_rQNM}
can be solved to find  the complex rQNM field value at the dipole location,
\begin{equation}
\tilde{\mathbf{f}}^{\rm r}_{\rm c}\left(\mathbf{r}_{0}\right)\cdot\mathbf{d}=\sqrt{\frac{\mathbf{d}\cdot\mathbf{G}\left(\mathbf{r}_{0},\mathbf{r}_{0};\omega_{\rm c}\right)\cdot\mathbf{d}}{A_{\rm c}\left(\omega_{\rm c}\right)}},\label{eq:fnut}
\end{equation}
where a real-valued dipole moment, $\mathbf{d}$, is assumed. The
above quantity is, in fact,  all one needs to perform an integration-free
normalization for the rQNM. Indeed, when inserted back into Eq.\,\eqref{eq:GF_rQNM}, one obtains  
\begin{equation}
\tilde{\mathbf{f}}^{\rm r}_{\rm c}\left(\mathbf{r}\right)=\frac{\mathbf{G}\left(\mathbf{r},\mathbf{r}_{0};\omega_{\rm c}\right)\cdot\mathbf{d}}{\sqrt{A_{\rm c}\left(\omega_{\rm c}\right)\left[\mathbf{d}\cdot\mathbf{G}\left(\mathbf{r}_{0},\mathbf{r}_{0};\omega_{\rm c}\right)\cdot\mathbf{d}\right]}}.\label{eq:QNM}
\end{equation}
Note that Eq.~\eqref{eq:QNM} now provides the full spatial profile
of the rQNM given that one also keeps track of the dipole response
at all other locations. However, in practice, the real part of the Green function is problematic for obtaining transverse system modes. \red{In general, the system Green function includes contributions from both transverse and longitudinal modes. In the presence of inhomogeneous and lossy media, these modes can be hard to separate \cite{Wubs1,Wubs2}, therefore,  solutions to Maxwell's equations subjected to dipole excitation can contain both types of modes, and  it is not clear how to obtain only the transverse modes, especially for lossy materials.} However, as a remedy to this problem, the (well behaved) imaginary part of the Green function at two different frequency points can be used to reconstruct the normalized transverse  field, as we discuss below.

We begin by finding the rQNM value at the dipole location, $\mathbf{r}_{0}$.
Consider Eq.\,\eqref{eq:GF_rQNM} at two different real frequencies, $\omega_{1}$
and $\omega_{2}$, that are, for example, located at either side of the rQNM resonance
frequency. By using the imaginary part of both sides for each equation
and following some simple algebra, we arrive at two independent
expressions for the real and imaginary parts of the complex rQNM, at the
dipole location:
\begin{align}
 & \mbox{Re}\left[\tilde{\mathbf{f}}^{\rm r}_{\rm c}\left(\mathbf{r}_{0}\right)\cdot\mathbf{d}\right]^2=\left\{ \text{Im}\left[\mathbf{d}\cdot\mathbf{G}\left(\mathbf{r}_{0},\mathbf{r}_{0};\omega_{1}\right)\cdot\mathbf{d}\right]\mbox{Re}\left[A_{\rm c}\left(\omega_{2}\right)\right]\right.\nonumber \\
 & \quad\quad\quad\left.-\text{Im}\left[\mathbf{d}\cdot\mathbf{G}\left(\mathbf{r}_{0},\mathbf{r}_{0};\omega_{2}\right)\cdot\mathbf{d}\right]\mbox{Re}\left[A_{\rm c}\left(\omega_{1}\right)\right]\right\} /B_{0},\label{eq:fnut1}
\end{align}
and
\begin{align}
 & \mbox{Im}\left[\tilde{\mathbf{f}}^{\rm r}_{\rm c}\left(\mathbf{r}_{0}\right)\cdot\mathbf{d}\right]^2=\left\{ \text{Im}\left[\mathbf{d}\cdot\mathbf{G}\left(\mathbf{r}_{0},\mathbf{r}_{0};\omega_{2}\right)\cdot\mathbf{d}\right]\mbox{Im}\left[A_{\rm c}\left(\omega_{1}\right)\right]\right.\nonumber \\
 & \quad\quad\quad\left.-\text{Im}\left[\mathbf{d}\cdot\mathbf{G}\left(\mathbf{r}_{0},\mathbf{r}_{0};\omega_{1}\right)\cdot\mathbf{d}\right]\mbox{Im}\left[A_{\rm c}\left(\omega_{2}\right)\right]\right\} /B_{0},\label{eq:fnut2}
\end{align}
where 
\begin{equation}
B_{0}=\text{Im}\left[A_{\rm c}\left(\omega_{1}\right)\right]\mbox{Re}\left[A_{\rm c}\left(\omega_{2}\right)\right]-\text{Im}\left[A_{\rm c}\left(\omega_{2}\right)\right]\mbox{Re}\left[A_{\rm c}\left(\omega_{1}\right)\right].
\end{equation}
Similarly, using the two space-point Green function, one requires the following additional set of two equations to obtain the normalized rQNM at all other locations away from the dipole, given the previously obtained $\tilde{\mathbf{f}}^{\rm r}_{\rm c}\left(\mathbf{r}_{0}\right)$:
\begin{align}
 & \mbox{Re}\left[\tilde{\mathbf{f}}^{\rm r}_{\rm c}\left(\mathbf{r}\right)\right]=\left\{ \text{Im}\left[\mathbf{G}\left(\mathbf{r}_{0},\mathbf{r};\omega_{1}\right)\cdot\mathbf{d}\right]\mbox{Re}\left[A_{\rm c}\left(\omega_{2}\right)\tilde{\mathbf{f}}^{\rm r}_{\rm c}\left(\mathbf{r}_{0}\right)\cdot\mathbf{d}\right]\right.\nonumber \\
 & \quad\left.-\text{Im}\left[\mathbf{G}\left(\mathbf{r}_{0},\mathbf{r};\omega_{2}\right)\cdot\mathbf{d}\right]\mbox{Re}\left[A_{\rm c}\left(\omega_{1}\right)\tilde{\mathbf{f}}^{\rm r}_{\rm c}\left(\mathbf{r}_{0}\right)\cdot\mathbf{d}\right]\right\} /B,\label{eq:felse1}
\end{align}
and
\begin{align}
 & \mbox{Im}\left[\tilde{\mathbf{f}}^{\rm r}_{\rm c}\left(\mathbf{r}\right)\right]\!=\!\left\{ \text{Im}\left[\mathbf{G}\left(\mathbf{r}_{0},\mathbf{r};\omega_{2}\right)\cdot\mathbf{d}\right]\mbox{Im}\left[A_{\rm c}\left(\omega_{1}\right)\tilde{\mathbf{f}}^{\rm r}_{\rm c}\left(\mathbf{r}_{0}\right)\cdot\mathbf{d}\right]\right.\nonumber \\
 & \quad\left.-\text{Im}\left[\mathbf{G}\left(\mathbf{r}_{0},\mathbf{r};\omega_{1}\right)\cdot\mathbf{d}\right]\mbox{Im}\left[A_{\rm c}\left(\omega_{2}\right)\tilde{\mathbf{f}}^{\rm r}_{\rm c}\left(\mathbf{r}_{0}\right)\cdot\mathbf{d}\right]\right\} /B, \label{eq:felse2}
\end{align}
where 
\begin{align}
 & B=\text{Im}\left[A_{\rm c}\left(\omega_{1}\right)\tilde{\mathbf{f}}^{\rm r}_{\rm c}\left(\mathbf{r}_{0}\right)\cdot\mathbf{d}\right]\mbox{Re}\left[A_{\rm c}\left(\omega_{2}\right)\tilde{\mathbf{f}}^{\rm r}_{\rm c}\left(\mathbf{r}_{0}\right)\cdot\mathbf{d}\right]\nonumber \\
 & \quad-\text{Im}\left[A_{\rm c}\left(\omega_{2}\right)\tilde{\mathbf{f}}^{\rm r}_{\rm c}\left(\mathbf{r}_{0}\right)\cdot\mathbf{d}\right]\mbox{Re}\left[A_{\rm c}\left(\omega_{1}\right)\tilde{\mathbf{f}}^{\rm r}_{\rm c}\left(\mathbf{r}_{0}\right)\cdot\mathbf{d}\right].\label{eq:b}
\end{align}
Note that, as can be seen from Eqs.~\eqref{eq:fnut1} and \eqref{eq:fnut2},
only the modal projection along the dipole direction is used to obtain
all modal components at all other locations through use of Eqs.~\eqref{eq:felse1}
and \eqref{eq:felse2}. It is worth mentioning that the technique presented here is found to be quite robust against the
chosen frequency values (within a maximum 5\% discrepancy for metal resonators and  1\% discrepancy for dielectric cavity systems).

\red{\subsection{FDTD implementation\label{sec:FDTDimplementation}}

As mentioned before, our proposed technique above  is quite general in its construction and can be applied, in principle, to any Maxwell solver, both in the time domain and in frequency domain. However, in this work, we will implement our method in FDTD, since it is  arguably one of the most popular Maxwell solvers among photonics and plasmonics communities. Indeed, the lack of efficient QNM calculation recipes in real time Maxwell solvers was one of the original motivations behind this work. To explain such motivation and the difficulties behind it in more detail, below we briefly review some of the previous works and attempts in dealing with FDTD\ Green functions and cavity mode calculations in the time domain.

Regarding the issue of obtaining the complex-frequency Green function in FDTD, there has been some work for studying Casimir effects \cite{Casimir1,Casimir2}, where a mathematically modified permittivity function is used to map the problem onto a complex frequency space, but it is limited to positive imaginary parts for the frequencies and therefore cannot be adopted to complex frequencies associated with QNMs.
In an earlier attempt to extract leaky mode behaviour in FDTD \cite{Yu2014},
a simple dipole-response normalization technique was proposed for
leaky photonic crystal cavities, but a normal mode picture was taken; indeed, a real-valued mode function was obtained that is known to lack the correct modal phase information (of an open cavity), and the method was cited to apply to dielectric structures only, for reasons that were not explained. The phase of the QNM is a necessity, e.g.,  for obtaining the correct Purcell factor as a function of position, particularly in plasmonics where very low quality factors are involved. A single dipole appoach with FDTD, with proper time windowing, can return the QNM, but, to allow a proper time windowing of the scattered field, this is usually restricted to dielectric cavities \cite{Kristensen2012} and is again is not in normalized form. 

There are other good reasons for  why implementing a dipole-response normalization of the QNMs in FDTD is so challenging. For example, a well known problem with using FDTD and other self-consistent Maxwell solvers,   stems from the
in-phase field component of finite-size dipoles, 
causing unwanted frequency shifts and a grid-size
dependence to the real part of the Green function
with equal space arguments, namely
${\bf G}({\bf r}_0,{\bf r}_0)$, a problem that also
occurs with  self-consistent local oscillators in FDTD, e.g., through the optical Bloch equations \cite{Ellen2017}.
The  Green function in FDTD can be directly obtained by using a point dipole source, defined through
\begin{equation}
\boldsymbol{\nabla}\times\boldsymbol{\nabla}\times{\mathbf{G}}\left(\mathbf{r},\mathbf{r}^{\prime};\omega\right)-k_0^2\varepsilon\left(\mathbf{r},\omega\right)\,{\mathbf{G}}\left(\mathbf{r},\mathbf{r}^{\prime};\omega\right)=k_0^2\mathbf{I}\delta(\mathbf{r}-\mathbf{r}^{\prime}),
\end{equation}
where $k_0^2=\omega^2/c^2$, $\varepsilon({\bf r},\omega)$ is the complex 
dielectric constant, and we assume non-magnetic materials.
For some applications, one can remove the grid-size dependence of FDTD dipoles, by subtracting off the solution from a homogeneous medium with the same computational gridding, so that the scattered Green function for an inhomogeneous medium
is  ${\bf G}^{\rm scatt}_{\rm FDTD}({\bf r}_0,{\bf r}_0)= {\bf G}^{\rm tot}_{\rm FDTD}({\bf r}_0,{\bf r}_0)-{\bf G}_{\rm FDTD}^{\rm hom}({\bf r}_0,{\bf r}_0)$. However, for the purpose of obtaining the transverse QNMs, particularly in the context of plasmonic resonators, the real part of the scattered Green function is not reliable, because the dipole response is contaminated by the influence of longitudinal modes.}

Before assessing the accuracy of this  FDTD
dipole technique for obtaining normalized rQNMs (and QNMs), we make a few remarks: (i) our normalization technique is quite easy to use with
any FDTD method, as it simply involves using the set of analytical
equations given by \eqref{eq:fnut1} to \eqref{eq:b}; moreover, the method is  practically instantaneous in time once the FDTD
dipole simulation is finished; (ii) because the normalization technique requires
no spatial information of the mode, from a practical perspective, the computational domain
termination using PML can be done as close as possible to the scattering
geometry before it alters the modal shape and eigenfrequency, resulting in significantly increased efficiency and less memory/run-time requirements.

\section{Applications to various cavity systems in nanophotonics and nanoplasmonics  \label{sec:Results}}

To demonstrate the reliability and capability of our normalization technique,
below we  consider five different cavity systems, including two hybrid cavity-waveguide
designs and a cavity-coupled photonic crystal waveguide. For all calculations, we use Lumerical FDTD \cite{lumerical} and a single mode performance over the frequency region of interest is assumed (and confirmed).

\subsection{Gold cuboids and dimers \label{subsec:metals}}

\begin{figure}
\begin{center}
\includegraphics[width=.99\columnwidth]{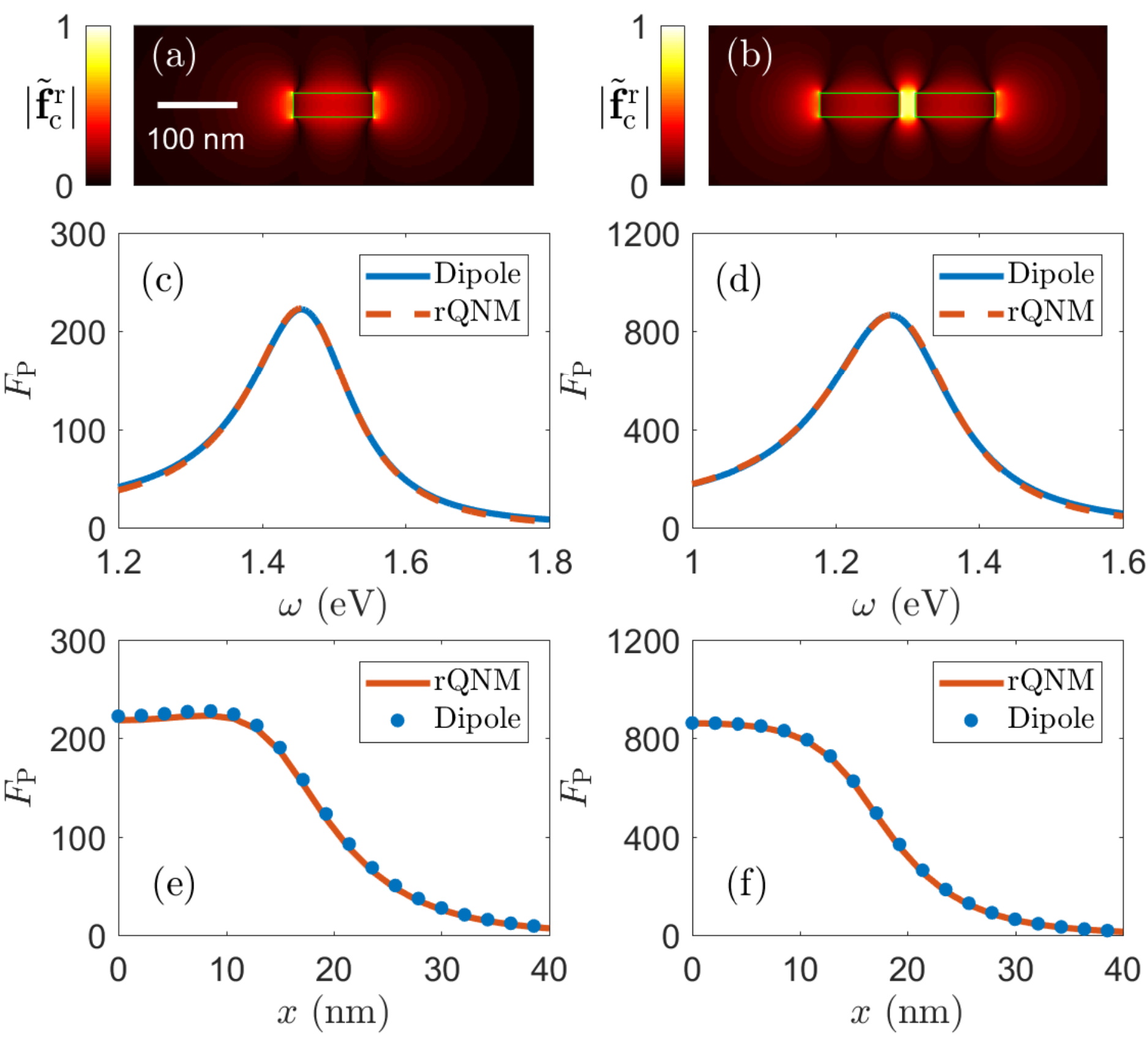}
\caption{(a-b) Computed rQNM spatial profile, $|\tilde{\mathbf{f}}^{\rm r}_{{\rm c}}|$ 
($y$-component), for single
and dimer cuboids, respectively. (c-d) Comparison between our dipole-normalized
rQNM calculation of the generalized Purcell factor, $F_{\rm P}$, against
fully vectorial calculations of Maxwell's equations. The dipole is located at 10 nm away from the surface along the nanorod axis for the single cuboid, and at the gap center for the dimer design. (e-f) Comparison of the rQNM technique in reproducing the position-dependent
 $F_{\rm P}$, again with full dipole calculations
in circles.\label{fig:Comparison-qnm}}
\end{center}
\end{figure}

First, we study a cuboid gold nanorod with dimensions of $30\times100\times30\,{\rm nm}^{3}$ placed in a homogeneous background with refractive index of $n_B=1.5$.
This acts as a single mode resonator over a wide range of frequencies
of more than $400\,{\rm meV}$, as shown in Fig.~\ref{fig:Comparison-qnm}.
For metallic regions, the dielectric function can be described using
the local Drude model,
\begin{equation}
\varepsilon_{{\rm metal}}\left(\mathbf{r},\omega\right)=1-\frac{\omega_{p}^{2}}{\omega\left(\omega+i\gamma_{p}\right)},
\end{equation}
where we use $\hbar\omega_{p}=8.29\,{\rm eV}$ and $\hbar\gamma_{p}=0.09\,{\rm eV}$
for the plasmon frequency and collision rate of gold \cite{Ge2014},
respectively. For the second example structure, two of the same nanorods
are used to form a dimer of gap spacing $h_{\rm g}=20\,{\rm nm}$. This forms
a plasmonic hot spot in the gap, but still behaves in a single mode manner over
a similar range of frequencies as the single nanorod.

As an  important  application of the mode technique, we study the spontaneous
emission enhancement factor for  dipole emitters when placed
nearby the resonant cavity structures of interest. Considering a quantum dipole
emitter polarized along $\mathbf{n}$, placed at position $\mathbf{r}$,
the generalized Purcell factor can be calculated using \cite{Anger2006}
\begin{equation}
F_{{\rm P}}({\bf r};\omega)=1+\frac{6\pi c^{3}}{\omega^{3}n_{B}}\,\mathbf{n}\cdot{\rm Im}\{\mathbf{G}\left(\mathbf{r},\mathbf{r};\omega\right)\}\cdot\mathbf{n}.
\end{equation}
For convenience of studying positions outside of metals (or outside the scattering geometry in general), we have added the extra factor of 1, which is derived from a Dyson equation scattering problem for dipole positions outside the resonator \cite{Ge2014njp}; otherwise
the Purcell factor should be defined without the extra factor of unity (e.g., for dipoles embedded within photonic crystal slabs).

For the two nanoresonator systems above, we use a computational domain
of $300\times500\times300\,\,{\rm nm}^{3}$ with a fine mesh of approximately $2\,{\rm nm}$
in every direction, terminated by PML in all directions.
The complex resonance frequencies  are found to be $\hbar\tilde{\omega}_{\rm c}=1.46 - 0.09i\,{\rm eV}$ and $\hbar\tilde{\omega}_{\rm c}=1.29 - 0.11i\,{\rm eV}$, respectively. The corresponding mode volumes at the dipole location \cite{Kristensen2012},
\begin{equation}
{\rm V_{eff}} = \frac{1}{{\rm Re}\{\varepsilon\left(\mathbf{r}_0\right)\tilde{\mathbf{f}}_{\rm c}^2\left(\mathbf{r}_0\right)\}},
\end{equation}
are also estimated to be ${\rm V_{eff}}/\lambda^3=8.5\times10^{-4}$ and ${\rm V_{eff}}/\lambda^3=1.6\times10^{-4}$.
As consequence of the integration-free nature of this technique, with
the small computational domain chosen, the entire simulation completes in
approximately 30 minutes on a standard desktop with
8 cores, even without spatial sub-meshing. It is of course useful to discuss how this approach compares with other QNM calculation techniques for general shaped cavity structures. First, in comparison to a previously
developed FDTD technique~\cite{Ge2014}, which uses a plane-wave
excitation with time filtering with the same sub-meshing: 
this simulation (for the same metal cavity) takes days to weeks to run as a much larger
simulation volume is required to carry out the spatial normalization procedure of the QNM. However, some additional time savings can be made by 
 implemented PML normalization~\cite{Sauvan2013} with FDTD~\cite{Kristensen2015}, but this requires one to use the PML data (often not available)\ and a volume integral with the electric and
magnetic fields (which requires additional care for field points inside the resonator). Having implemented all three approaches in FDTD, we find that our newly presented dipole normalization method is easily the most efficient to work with, and the most powerful (e.g., it can also do periodic background media as we demonstrate later). 
With regards to the frequency-domain dipole approach using complex frequencies in COMSOL~\cite{Bai2013}, we have found that for a complete analysis, this approach takes about the same computational time as the proposed dipole FDTD method, but only for the smaller spatial domains, such as with the nanoresonator devices. However, in our experience,
larger hybrid devices run into extreme memory requirements when using finite-element
solvers (e.g. with  COMSOL). Both approaches have their strengths an weaknesses though, and the COMSOl approach is better for solving multiple overlapping modes (given sufficient computational resources).

For the metal resonator calculations here, 
we implemented a numerical dipole source,
located  $10\,{\rm nm}$ away from the metallic surface along the
$z$-axis in the case of single cuboid, and at the center of the gap
in the case of the cuboid dimer. One can use the same spatial dipole position to perform both full numerical
Purcell factor calculations as well as the rQNM calculation, all within
the same one-time simulation, but for different dipole positions, we stress that there is no need to recalculate the rQNM. In Fig.~\eqref{fig:Comparison-qnm},
we  plot the computed mode profile of the single rQNM of interest for
each case along with the  rQNM-calculated Purcell factor (which is analytic, after obtaining the mode numerically of course)\ in dashed-red
that compares very well with full dipole calculations using Eq.~\eqref{eq:GFnumeric}
in solid-blue (the accuracy is within a few \%).

To more rigorously confirm the reliability of our rQNM technique, we
next perform position-dependent Purcell factor studies for both of
the structures discussed above. In Fig.~\ref{fig:Comparison-qnm},
each circle shows an independent dipole calculation (e.g., with no
approximations) done at a particular position, while the solid line
is calculated using the same rQNM calculated before (which only has to be computed once). In both cases
we move away along the $x$-direction up to $40\,{\rm nm,}$ where an excellent
agreement between our semi-analytical results and the full numerical
results is achieved in all cases. In particular, note that the Purcell
factor behavior before reaching $x=10\,{\rm nm}$  is qualitatively different
for the single cuboid and the dimer resonator, and clearly the rQNM calculation
accurately captures both trends. Such a good level of agreement can be in principle obtained
at every location around the resonator for which the rQNM expansion
remains valid. At  distances far from the resonator, as  discussed below in subsection \ref{subsec:reg}, our integration-free rQNMs also recovers the correct physical behavior, and thus we speculate that this rQNM picture can be used in both near and far fields from the resonator, with practically no distinction to the QNM in the near field---where the rQNM and QNM are identical.

\subsection{Implicit far-field regularization:  regularized QNMs vs divergent QNMs \label{subsec:reg}}

In the above, we have shown how the rQNMs are practically identical to the QNMs in the near field (and certainly within the resonator), which is a consequence of the single mode approximation working very well. Unfortunately, in the far field, the single mode approximation of a cavity mode must fail. Although it may seem like an academic question, it is important to have physically meaningful fields at these far field locations as well, e.g., to compute the field that would be detected (e.g., by a detector)\ from the resonators in quantum optics \cite{Ge2015prb}. In this section, we demonstrate the ``regularized'' nature of our rQNMs  when going to the far-field space region. 

\begin{figure}
\begin{center}
\includegraphics[width=.99\columnwidth]{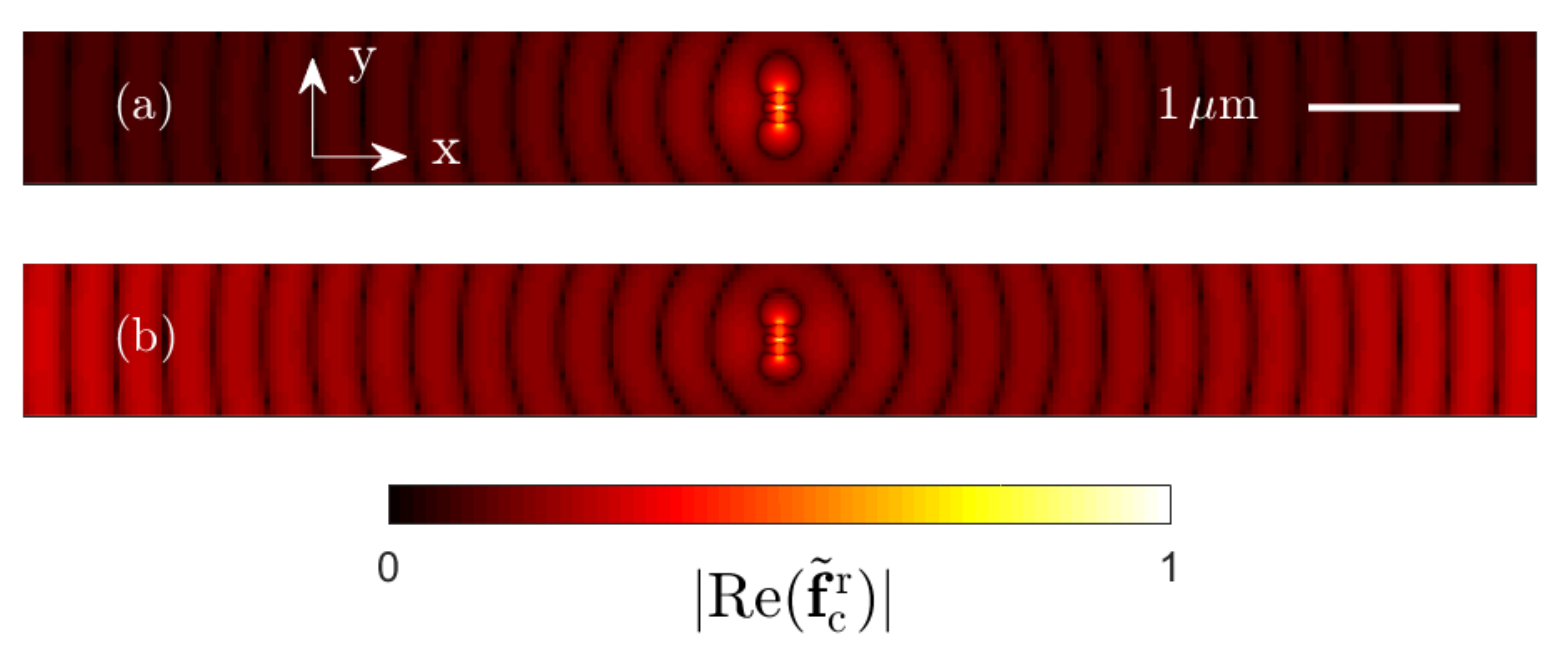}
\vspace{.1cm}
\includegraphics[width=.99\columnwidth]{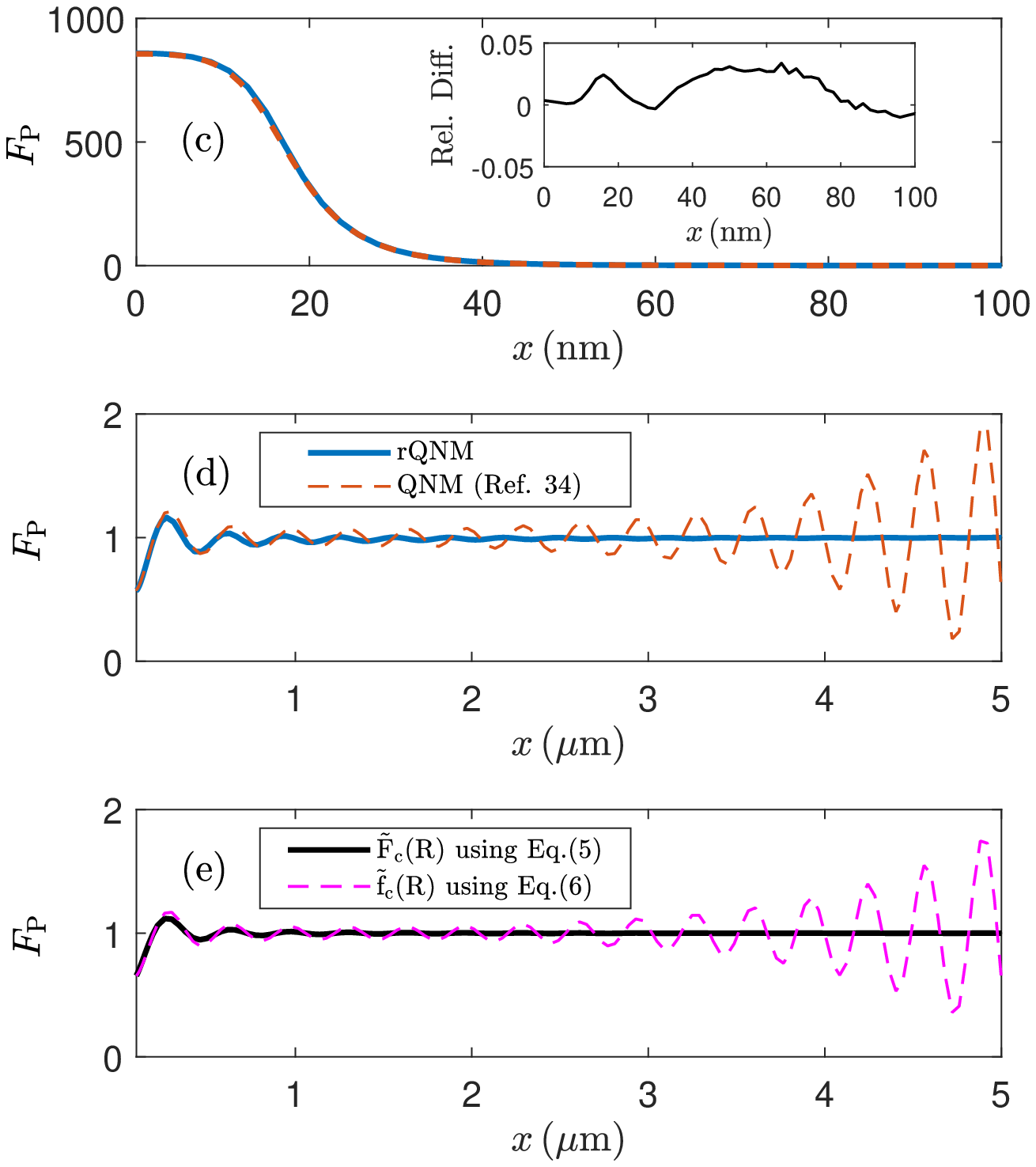}
\caption{\red{Near-field and far-field behavior of the integration-free rQNM versus QNM for the cuboid gold dimer (note the latter is calculated using a completely different FDTD technique~\cite{Ge2014}). (a-b) Extended 2D map of the $|{\rm Re}(\tilde{\mathbf{f}}^{\rm r}_{\rm c}(\mathbf{r}))|$ and $|{\rm Re}(\tilde{\mathbf{f}}_{\rm c}(\mathbf{r}))|$, respectively. Each colormap is individually normalized to one and the same nonlinear scaling is used to better show the differences. (c-d) Comparison of the corresponding Purcell factors, calculated in the near-field region and in the far-field region, respectively. The inset in (c) shows the relative difference between these two independent calculations. (e) Purcell factor calculations based on the Dyson equation treatment of 
$\mathbf{F}_{\rm c}(\mathbf{r})$ using Eq.\,\eqref{eq:bigF} in real frequency space, and $\tilde{\mathbf{f}}_{\rm c}(\mathbf{r})$ using Eq.\,\eqref{eq:bigFcmplx} in complex frequency space. In all of these plots, $x=0$ represents the origin placed at the dimer gap center.}\label{fig:divg}}
\end{center}
\end{figure}

As mentioned before, the system QNMs are solutions to a non-Hermitian Maxwell problems with open boundary conditions and are associated with complex resonant frequencies or complex wavevectors. \red{The negative imaginary part for the complex frequency, that describes the energy decay in time domain, leads to an exponential growth of the QNMs in space through $e^{(\gamma_c n_B/c) x}$, as one gets far enough away from the resonator. As a general rule of thumb, the divergence behavior of the QNM is  become significant around $x_{\rm divg}$ such that $(\gamma_c n_B/c)x_{\rm divg}\approx 1$. For example, using the imaginary part of the complex frequency for the cuboid dimer design,
where $\gamma_c \approx 0.11~$eV,  $x_{\rm divg}=1-2\,{\rm \mu m}$ is estimated using this simple argument, which agrees with what will be discussed shortly in Fig.\,\ref{fig:divg}.} This known feature has been one of the main challenges in normalizing QNMs, and have raised questions as to whether these modes can be used to properly describe certain aspects of experiments in the far-field and with input-output formalisms.  

As highlighted earlier, a Dyson equation approach can be used to regularize the divergent QNM in the far-field through use of Eq.\,\eqref{eq:bigF} in real frequency space. But this approach also requires an additional spatial integration step and is rather complicated for metal structures to implement. However, the computed rQNM, $\mathbf{f}^{\rm r}_{\rm c}(\mathbf{r})$, is obtained in real frequency space and is thus already regularized. To demonstrate this  regularization behavior, we next investigate the positional dependence of the generalized Purcell factor for dipole emitters coupled to the cuboid dimer structure, both in the near field and far field  regimes. \red{We first show extended 2D maps ($10\,{\rm \mu m}$ along $x$-axis) of the rQNM and QNM spatial modes (as calculated in Ref.~\cite{Ge2014}), respectively in Fig.~~\ref{fig:divg}(a-b). A nonlinear color scaling is used to enhance the differences between the two approches. In particular, in the far-field the QNM diverges where as the rQNM does not, because our real-frequency QNM calculation technique captures the proper (sum over modes)\ propagation effects to the far-field. To better highlight these differences, we plot the Purcell factor at various positions using both rQNM and QNM, also in Fig.~\ref{fig:divg}.} In Fig.~\ref{fig:divg}(c), for positions close to the resonator, our rQNM Purcelll factor gives an excellent agreement with the QNM as calculated based on the recipe given in Ref.~\onlinecite{Ge2014}. However, in the far-field region, as shown in Fig.~\ref{fig:divg}(d), the rQNM behaves in a converged manner in comparison to the  QNM. In particular, it follows the prediction of the post-calculated regularized QNM using the Dyson integral of Eq.\,\eqref{eq:bigF}, that is shown in Fig.~\ref{fig:divg}(e). For completeness however, if one needs to obtain the true divergent behavior of the system QNM in the far-field using our integration free approach, one can easily use the complex-frequency Dyson treatment of Eq.\,\eqref{eq:bigFcmplx}; since, to a very good approximation,  our rQNM is equivalent to the system QNM inside (and near) the resonator, as clearly demonstrated in in Fig.~\ref{fig:divg}(e). Given the small value of the Purcell factor in the far-field (note we are zooming in from 1000 to 1), the minor discrepancies between (d) and (e) are attributed to small numerical errors (from the spatial integration primarily), and are not a general concern. Thus, these rQNM can be practically used at all spatial positions, without any spatial integration at all. If one desires the rQNM outside the simulation volume domain, then one can easily obtain either the QNM or the rQNM using the Dyson equation as shown above. Thus there is no need to simulate large regions of homogeneous space outside the scattering geometry, and for continuous waveguide one can also safely use PML to limit the space- and run-time requirements.

\subsection{Hybrid metal-dielectric systems \label{subsec:hybrids}}

To further emphasize the generality of our technique, and show
its applicability for use in more complex emerging devices
in hybrid plasmonics, we next study a set of two hybrid
devices where the plasmonic single cuboid resonator is either embedded
in a ridge waveguide or is placed inside a groove waveguide, such that the short-range confined mode of the cuboid is coupled to the long-range propagating mode of the waveguide. The dielectric beam waveguide
is made of silicon-nitride, with a refractive index of $n_{{\rm diel}}=2.04$
and dimensions of $600\times800\times6000\,\,{\rm nm}^{3}$. One motivation
behind such devices is to design transmission drop lines, but here,
for consistency with our other investigations, we focus on the far-field
collection efficiency of dipole emitters coupled to the plasmonic
sub-system, that can be also of interest in the design of integrated broadband single photon sources; for example, one could excite embedded quantum dots incoherently, and monitor the output scattered field of a single exciton state. For both of these cases, a computational domain of $10\times5\times5\,\,{\rm \mu m}^{3}$
is used to ensure a sufficiently long waveguide terminated by PML in all directions. Similar to before, a fine mesh of
approximately $2\,{\rm nm}$ was used over the metallic region, that
was then refined into the courser mesh of $40\,{\rm nm}$ everywhere else,
using the non-uniform meshing technology in Lumerical FDTD~\cite{lumerical}. 

\begin{figure}
\begin{center}
\includegraphics[width=.99\columnwidth]{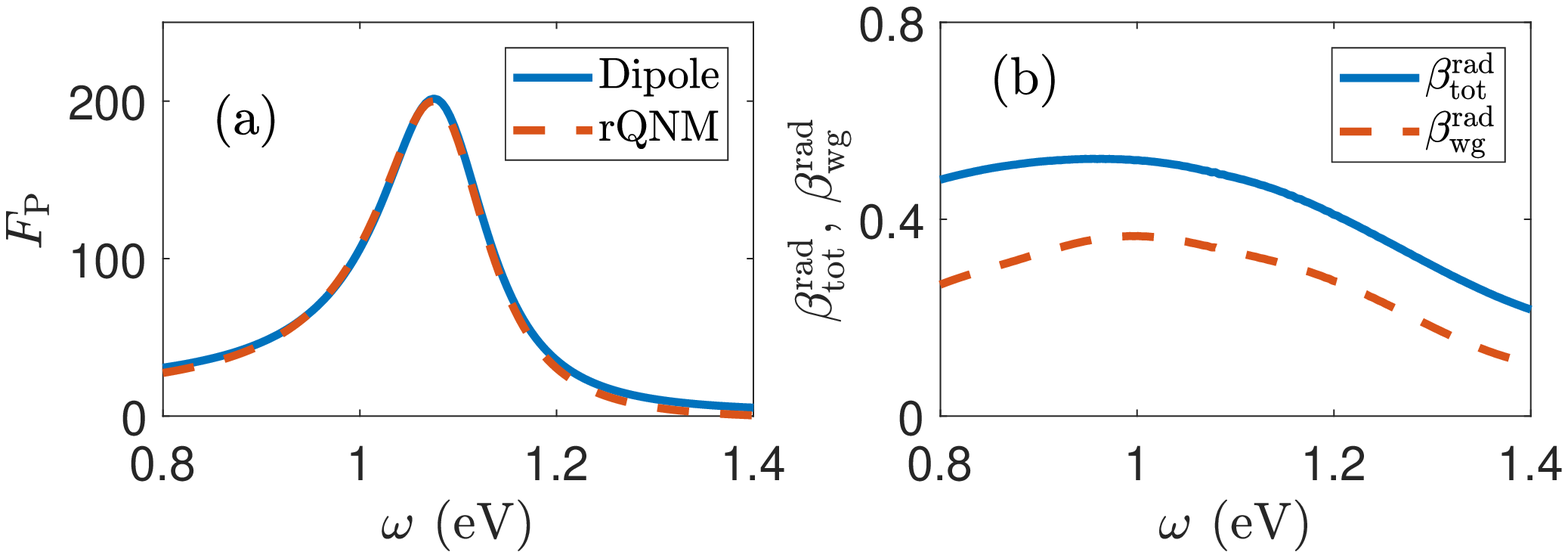}
\vspace{.2cm}

\includegraphics[width=.99\columnwidth]{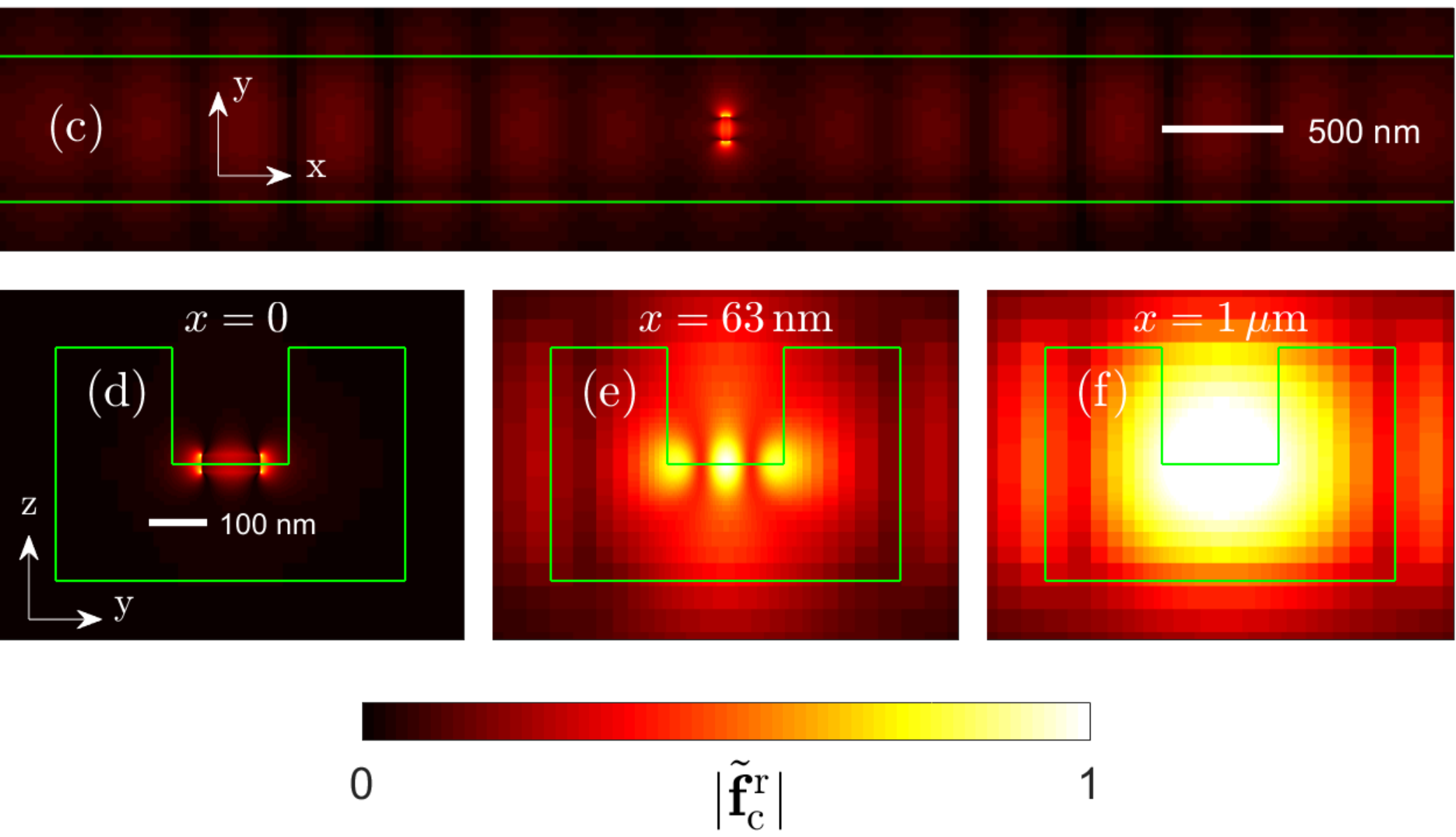}
\caption{(a) Comparison between our dipole-normalized rQNM calculation of the
generalized Purcell factor, $F_{\rm P}$, against fully vectorial calculations of Maxwell's equations, for the gold single cuboid coupled to the
ridge waveguide. The dipole emitter is placed at 10 nm away from the nanorod surface along the $y$ axis. (b) Radiative beta factors: the $\beta^{\rm rad}_{\rm tot}$ that quantifies the
entire far-field radiation as well as the $\beta^{\rm rad}_{\rm wg}$ that quantifies
the far-field radiation confined within the waveguide. (c) An 
$xy$ cut of the calculated rQNM spatial profile at $z=0$ and (d-f) show three different $yz$ cuts as labeled, to show the technique ability
in capturing modal details. \red{Each colormap is individually normalized to one and a nonlinear scaling for improved visualization is used in (c)}.\label{fig:hybrid-pf}}
\end{center}
\end{figure}

\begin{figure}[t!]
\begin{center}
\includegraphics[width=.99\columnwidth]{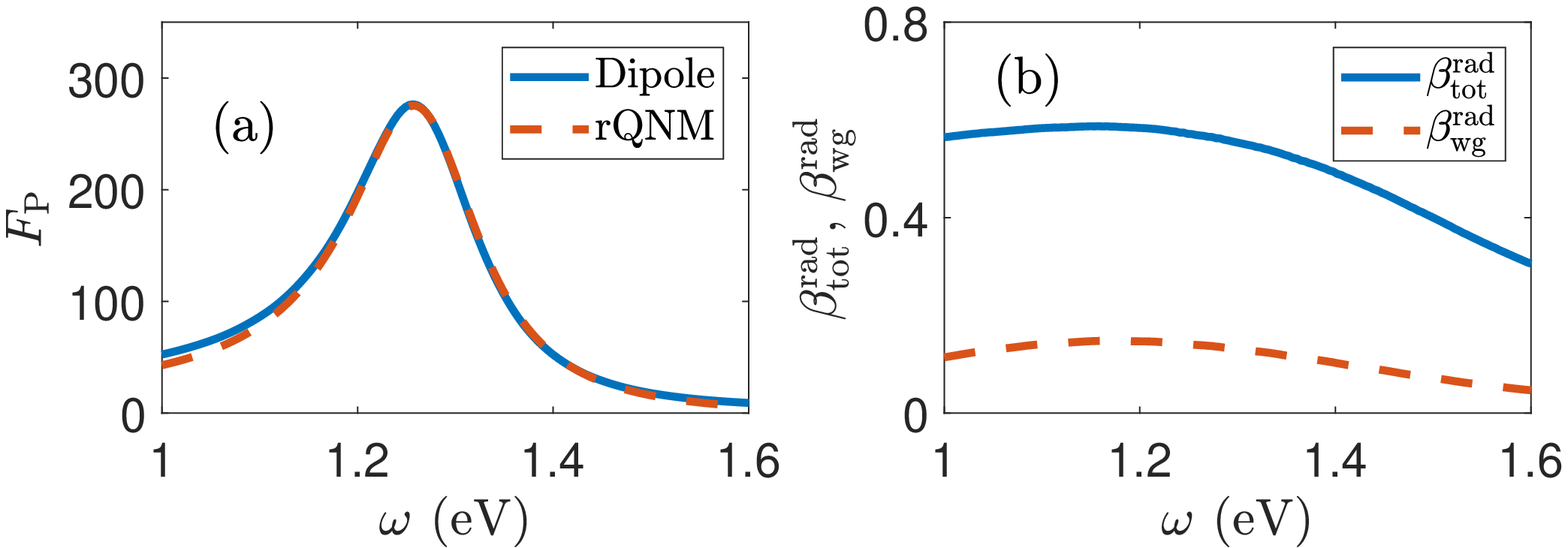}
\vspace{.2cm}

\includegraphics[width=.99\columnwidth]{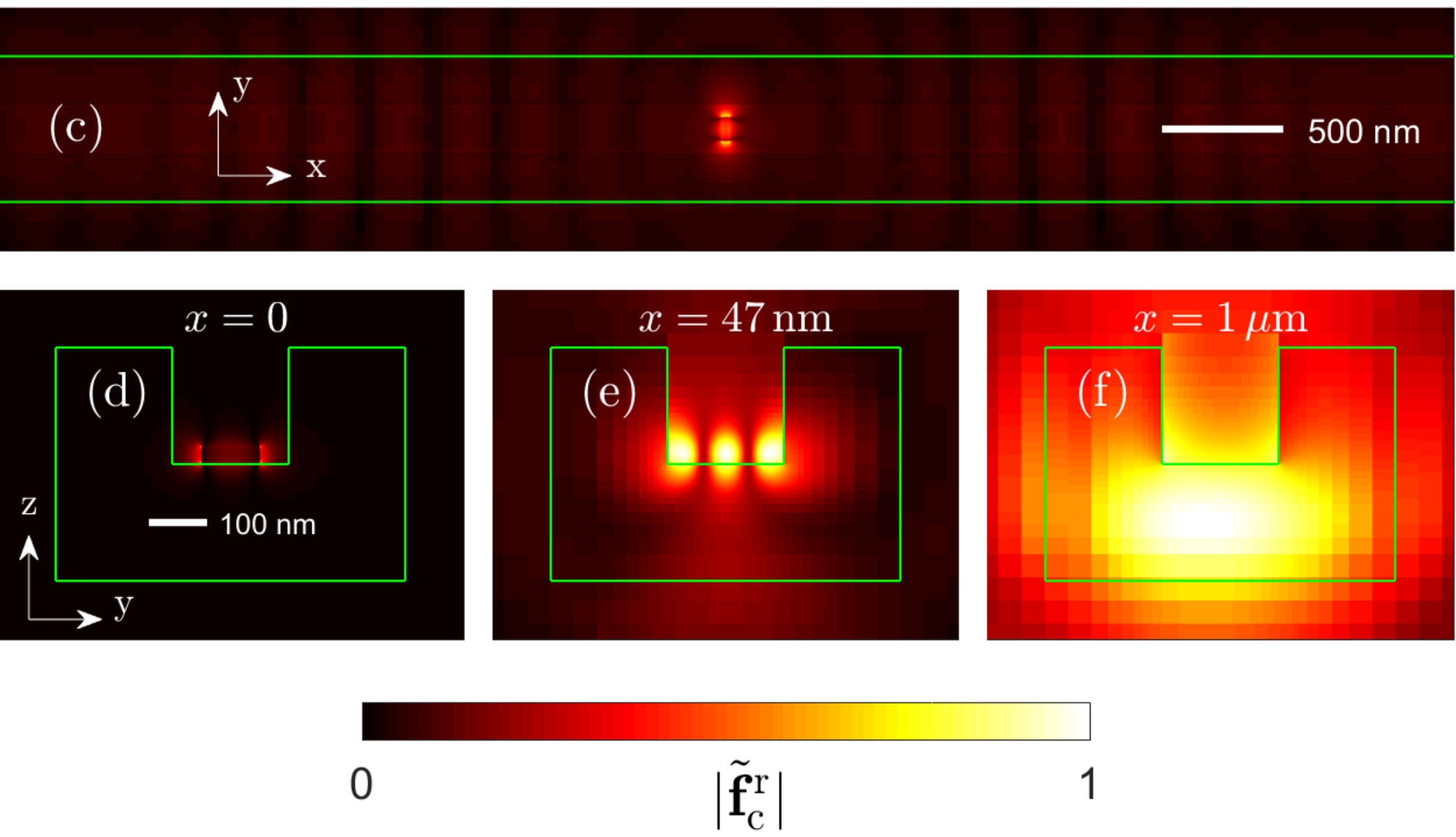}
\caption{Panels represent  the same quantities plotted in Fig.\,\ref{fig:hybrid-pf} but for the single nanorod placed in the groove waveguide.\label{fig:hybrid-pf-2}}
\end{center}
\end{figure}

In Fig.\,\ref{fig:hybrid-pf}(a)
and Fig.\,\ref{fig:hybrid-pf-2}(a), we  plot the generalized
Purcell factor, $F_{\rm P}$, of a dipole emitter placed $10\,{\rm nm}$
away from the metallic surface (see also Fig.\,\ref{fig:schematic}). The complex resonance frequencies extracted are
$\hbar\tilde{\omega}_{\rm c}=1.08 -0.07i\,{\rm eV}$ and $\hbar\tilde{\omega}_{\rm c}=1.26 - 0.08i\,{\rm eV}$, respectively. The corresponding mode volumes are also estimated to be ${\rm V_{eff}}/\lambda^3=6.5\times10^{-4}$ and ${\rm V_{eff}}/\lambda^3=6.3\times10^{-4}$.
Both full-dipole calculations (in solid-blue) and rQNM calculations in (dashed-red)
are shown, once again with a  good agreement. In comparison to isolated resonators,
there is a small discrepancy between the rQNM and full dipole calculations at frequencies very far from the resonance, which
is likely attributed to the non-negligible influence from dielectric wall/surface scattering effects (possibly yielding small background modes at other frequencies) and the fact that now the single resonator interacts with a  propagating mode of the waveguide. Similar to before, a single mode performance is still achieved, but the resonance is red-shifted and sits lower in terms of maximum emission enhancement, due to the lower index mismatch provided by the waveguide.


With plasmonics, one always has some optical quenching of the mode, and it is also important to know how much of the radiative emission of a quantum emitter can couple
to the target waveguide mode in the two different configurations mentioned above.
This can be quantified for the radiation emitted into the waveguide, $\Gamma_{\rm wg}^{\rm rad}$, using the waveguide radiative ``beta factor''
\begin{equation}
\beta_{\rm wg}^{\rm rad} = \frac{\Gamma_{\rm wg}^{\rm rad}}{\Gamma_{\rm P}},
\end{equation}
and for the total radiation available in the entire $4\pi$ far-field space (away from the resonator), $\Gamma_{\rm tot}^{\rm rad}$, using
\begin{equation}
\beta_{\rm tot}^{\rm rad} = \frac{\Gamma_{\rm tot}^{\rm rad}}{\Gamma_{\rm P}},
\end{equation}
where $\Gamma_{\rm P}=F_{\rm P}\Gamma_{\rm 0}$, and $\Gamma_0$ the spontaneous emission rate in a homogeneous medium (free space or the dielectric).
Thus the total nonradiative coupling is simply 
\begin{equation}
\beta^{\rm nrad} = 1- \beta_{\rm tot}^{\rm rad}.
\end{equation}

The radiative beta factors are shown in Fig.\,\ref{fig:hybrid-pf}(b) and Fig.\,\ref{fig:hybrid-pf-2}(b),
where the solid-blue shows the total radiation available in the far-field,
and the dashed-red shows only its portion transmitted through the waveguide
interface. This factor also gives the quantum efficiency of a dipole emitter. It is seen that, depending on the waveguide design, the far-field emission can be quite different.
In particular, the total far-field radiation from the dipole for the groove waveguide is higher than in the ridge waveguide over its frequency band, mainly because the nanoparticle is not embedded in the dielectric region.
However, the dipole emission within the waveguide is considerably less for the groove design, again because the nanoparticle is not embedded in the dielectric. Moreover, the fact that close to 40\% of dipole emission can be detected at the end of the ridge waveguide, offers some benefits in comparison to all-metallic plasmonic waveguides (which introduced additional waveguide losses), in applications where signal strength (brightness) from single emitters is more critical.

To better see the mode pattern in the waveguide, in Fig.~\ref{fig:hybrid-pf} and Fig.~\ref{fig:hybrid-pf-2},
we have plotted four different surface slices of the calculated rQNM. In 
Figs.~\ref{fig:hybrid-pf}-\ref{fig:hybrid-pf-2} (c), we show an $xy$-cut at the beam
center such that the top view of the whole system is shown. This shows
a clear pattern of the complex rQNM that has features from both the
plasmonic resonator and the nanobeam waveguide. In Figs.~\ref{fig:hybrid-pf}-\ref{fig:hybrid-pf-2} (d), an $yz$-cut at
the center of the plasmonic cuboid where the dominant plasmonic behavior
is shown. Finally, in Figs.~\ref{fig:hybrid-pf}-\ref{fig:hybrid-pf-2} (e-f) two other $yz$-cuts at distant locations from the single cuboid are shown,
where the transition form the localized plasmonic mode to the propagating
waveguide mode is y observed.

\subsection{Photonic crystal slab coupled cavity-waveguide \label{subsec:PC}}

For our final example, we apply our technique to a class of devices in photonic crystal platforms, namely cavities coupled to periodic waveguides, that are used for applications  such as single photon sources and channel drop filters. This not only demonstrates the reliability of our technique for use in dielectric systems with large quality factors, but also tackles the very difficult problem of normalizing QNMs for the localized cavities that are subjected to the Bloch periodic propagation of the waveguide, as the dominant outgoing channel \cite{KristensenOL2014,Malhotra2016}; in this regime, there are additional complexities and difficulties encountered to regularize an infinite spatial integration \cite{KristensenOL2014,Malhotra2016}.

\begin{figure}[ht]
\begin{center}
\includegraphics[width=.99\columnwidth]{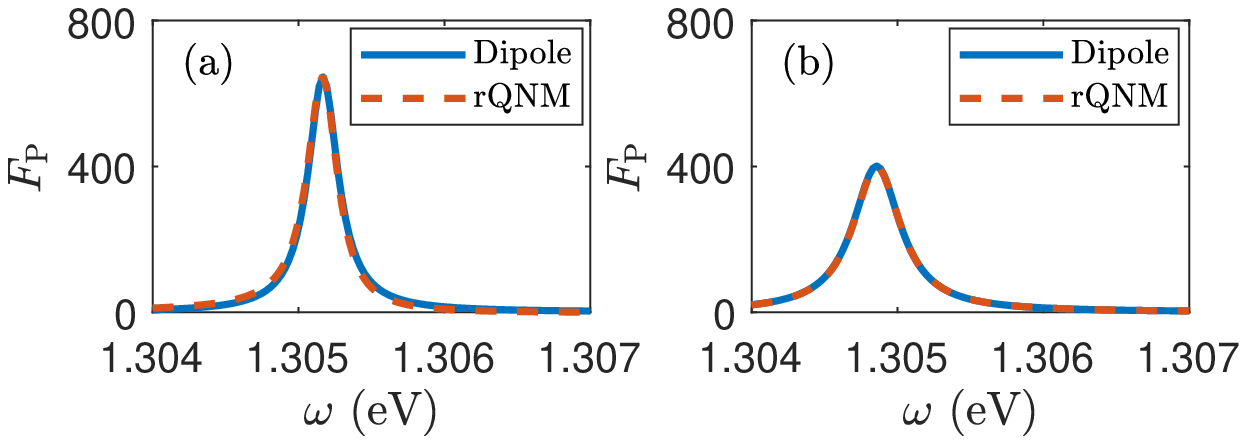}
\includegraphics[width=.99\columnwidth]{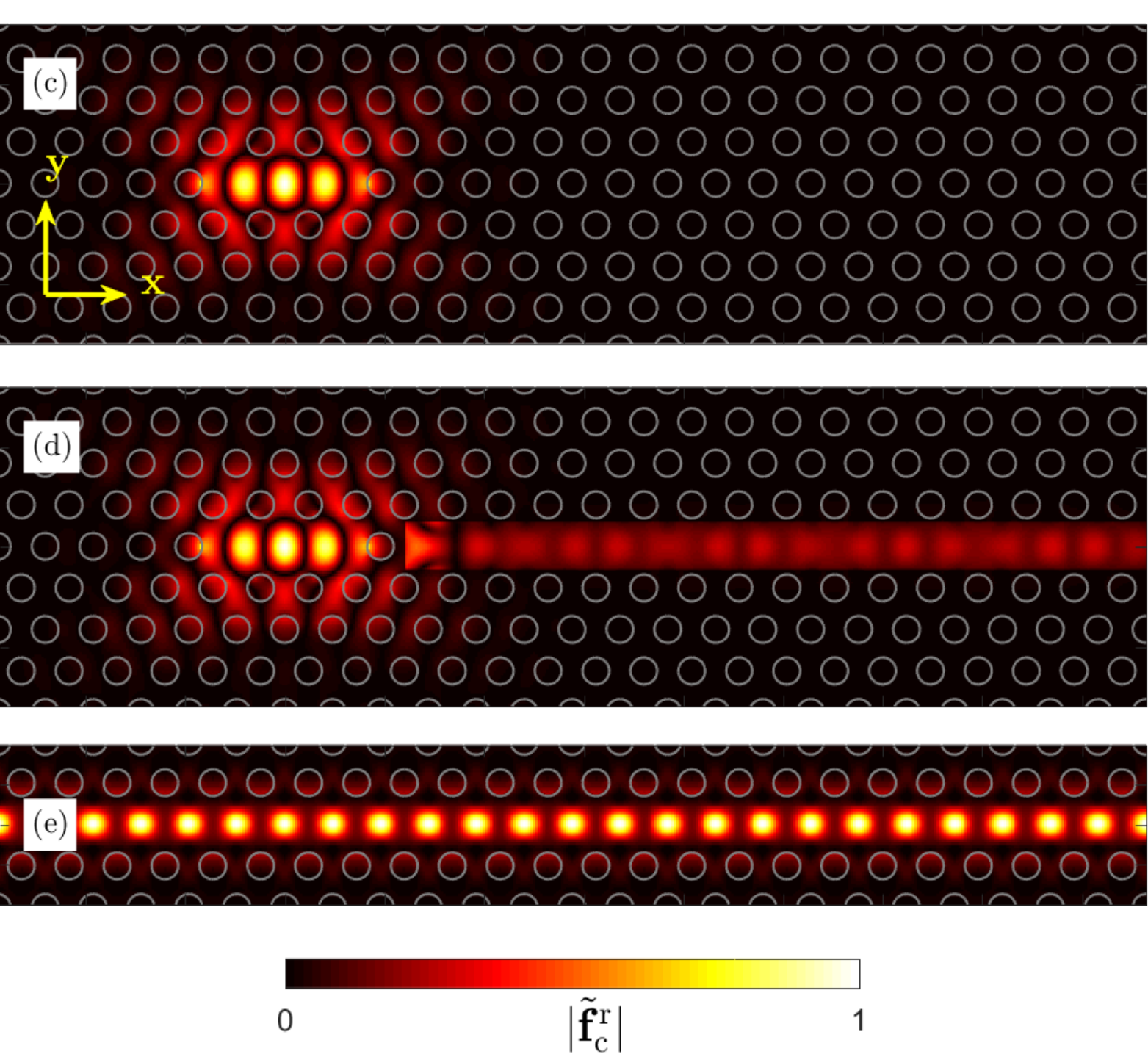}
\caption{(a-b) Comparison between our dipole-normalized rQNM calculation of the generalized Purcell factor, $F_{\rm P}$, against full-dipole vectorial calculations of Maxwell's equations, for the isolated L3 cavity and the L3 cavity side-coupled to the waveguide, respectively. The dipole emitter is placed at center of the cavity and polarized along the $y$-axis. (c-d) An $xy$ cut of the calculated rQNM spatial profile at the center of the slab where in (d) the waveguide portion is nonlinearly enhanced for better visualization. \red{ (e) The propagating Bloch mode of a section of the infinite waveguide for comparison (see inside text for details). Each colormap is individually normalized to one.}\label{fig:l3w1}}
\end{center}
\end{figure}

We use a triangular photonic crystal slab of refractive index of $n=3.5$, where the lattice constant is $d=240\,{\rm nm}$, the hole radius is $r=0.28\,d$ and the slab thickness is $h=164\,{\rm nm}$. The structure is $9\times25\,d$ in size, and is finely meshed with 20 points per lattice period. In particular, we consider two devices, the isolated L3 cavity on its own, and the L3 cavity side-coupled to a W1 waveguide. To minimize any relevant numerical discrepancies, the L3 cavity is placed to the side such that exactly the same structure and computational domain is used for both devices; see Fig.\,\ref{fig:l3w1}.

Depicted in Figs.\,\ref{fig:l3w1}(a)-(b), as for the other cavity structures, we first confirm a very good agreement between the  QNM-calculated Purcell factor and the fully vectorial dipole calculations; here we consider a dipole emitter placed inside the L3 cavity, at the antinode and aligned along the $y$-axis, according to the coordinate system shown in the schematic of Fig.\,\ref{fig:schematic}. As seen, introducing the waveguide reduces the peak enhancement by close to a factor of 2. The resonance frequencies (real part) are computed to be $\hbar{\omega}_{\rm c}^{\rm L3}=1.3052\,{\rm eV}$ and $\hbar{\omega}_{\rm c}^{\rm L3W1}=1.3049\,{\rm eV}$, with the corresponding quality factors of $Q^{\rm L3}=5,200$ and $Q^{\rm L3W1}=3,300$. The associated mode-volumes, calculated at the modal antinode at the center of the L3 cavity are also ${\rm V_{eff}^{L3}}/\lambda^3=0.0142$ and ${\rm V_{eff}^{L3W1}}/\lambda^3=0.0145$, which are found to be very similar for this particular structure (as is often assumed in the community without proof, but in general this may not be the case, especially for low-$Q$ cavities).

In  Figs.~\ref{fig:l3w1}(c) and \ref{fig:l3w1}(d), we  display the rQNM spatial profile as calculated for both photonic crystal devices, where the in-waveguide section for the coupled device is enhanced for visualization. These two represent $xy$ cuts at the center of the slab. It is evident that the rQNM inside the waveguide behaves  different than the waveguide Bloch mode and somewhat inherits the shape of the L3 mode, with repetitions occurring due to the phase dependence of the cavity QNM~\cite{Kristensen:17}. Therefore, the naive assumption that simply the system behavior inside the waveguide follows the usual W1 propagation, may not be taken. In addition, note that the details of the transition for the system mode to go from a localized mode within the cavity to propagating within the waveguide, that is nontrivial, is also well captured by the calculated rQNM.

\section{Conclusions\label{sec:Conclusions}}

In summary, we have introduced a new normalization technique for obtaining  
regularized QNMs  of leaky optical cavities and plasmonic resonators, and implemented it with the widely used FDTD algorithm in real frequency space. Our  technique requires
no spatial integration for post processing normalization of the QNMs, but rather a set of easy-to-use
analytical equations are provided that only require the self-consistent
response to a dipole excitation at two different frequencies, essentially exploiting an inverse Green function approach. We find
this  technique to be extremely efficient on both computational memory and run time requirements, and easy to use.
We exemplified this dipole normalization technique for several different
arbitrarily-shaped plasmonic resonators, dielectric photonic crystal slabs, and hybrid devices in order to show its generality and applicability to a wide range of nanophotonic systems.
In particular, the spontaneous emission enhancement factor was studied for quantum emitters placed nearby these systems, where a very good agreement (within a few~\% at the desired
frequency range) between our rQNM calculation and
fully numerical solutions of Maxwell's equations were obtained. Moreover, this new technique requires no further regularization of the rQNM in the far-field, as it readily returns the non-divergent system response at distances far away from the resonator. Since the method is easy to implement in commonly used FDTD solvers such as those available from Lumerical solutions \cite{lumerical} and MEEP \cite{meep}, it is an attractive tool for the community seeking true ``cavity modes'' for a wide range of complex  structures. 

\section*{Acknowledgments}

We thank  Queen\textquoteright s University and the Natural Sciences and Engineering Research Council of Canada for financial support. We also thank Lumerical Solutions Inc. for support, and Philip Kristensen and Simon Axelrod for excellent suggestions and discussions.

\bibliographystyle{apsrev4-1}
\bibliography{Refs}

\end{document}